\documentclass{elsart}
\usepackage{epsfig}
\usepackage{color}
\journal{Nucl. Instrum. Methods A}
\usepackage{amssymb}

\begin{document}

\begin{frontmatter}

% Title, authors and addresses
\title{Measurement of the specific activity of $^{39}$Ar in natural 
argon}

% use optional labels to link authors explicitly to addresses:
% \author[label1,label2]{}
% \address[label1]{}
% \address[label2]{}

%\author[AQ]{\it{AUTHORS}}
\author[PV]{P.~Benetti},
\author[Prin]{F.~Calaprice}, 
\author[PV]{E.~Calligarich}, 
\author[PV]{M.~Cambiaghi},
\author[NA]{F.~Carbonara},
\author[AQ]{F.~Cavanna}, 
\author[NA]{A.~G.~Cocco},
\author[AQ]{F.~Di Pompeo},
\author[LNGS]{N.~Ferrari\thanksref{deceased}},
\author[NA]{G.~Fiorillo},
\author[Prin]{C.~Galbiati},
\author[PV]{L.~Grandi},
\author[NA]{G.~Mangano},
\author[PV]{C.~Montanari},
\author[LNGS]{L.~Pandola\corauthref{cor}},
\ead{pandola@lngs.infn.it}
\corauth[cor]{INFN, Laboratori Nazionali del Gran Sasso,
S.S. 17 bis km 18+910, I-67100 Assergi (AQ), Italy.
Telephone number: +39 0862 437532, Fax number: +39 0862 437570}
\author[PV]{A.~Rappoldi},
\author[PV]{G.~L.~Raselli},
\author[PV]{M.~Roncadelli},
\author[PV]{M.~Rossella},
\author[PV]{C.~Rubbia},
\author[NA]{R.~Santorelli},
\author[Krak]{A.~M.~Szelc},
\author[PV]{C.~Vignoli}
\and \author[Prin]{Y.~Zhao}

%\address[AQ]{Addresses}
\address[PV]{INFN and Department of Physics, University of Pavia, Pavia, Italy}
\address[Prin]{Department of Physics, Princeton University,  
Princeton NJ, USA}
\address[NA] {INFN and Department of Physics, University of Napoli 
``Federico II'', Napoli, Italy}
\address[AQ] {INFN and Department of Physics, University of L'Aquila, 
L'Aquila, Italy}
\address[LNGS] {INFN, Gran Sasso National Laboratory, Assergi, Italy} 
\address[Krak] {Institute of Nuclear Physics PAN, Krak\'ow, Poland}
\thanks[deceased]{Deceased}

\begin{abstract}
% Text of abstract
We report on the measurement of the specific activity of $^{39}$Ar in 
natural argon. 
The measurement was performed with a 2.3-liter two-phase (liquid and gas) argon 
drift chamber. The detector was developed by the WARP Collaboration as a prototype 
detector for WIMP Dark Matter searches with argon as a target.
The detector was operated for more than two years at 
Laboratori Nazionali
del Gran Sasso, Italy, at a depth of 3,400\,m w.e. 
The specific activity measured for $^{39}$Ar 
is 1.01$\pm$0.02(stat)$\pm$0.08(syst)\,Bq per kg of $^{\rm nat}$Ar.
\end{abstract}


\end{frontmatter}

% main text
\section{Introduction} \label{sec1}
A 2.3-liter two-phase (liquid and gas) argon drift chamber~\cite{warp-prototype} was 
developed and built by the WARP collaboration as a prototype detector for WIMP Dark 
Matter searches with argon as a target~\cite{warp-proposal}.
The detector was operated for more than two years by the WARP collaboration at Laboratori 
Nazionali
del Gran Sasso, Italy, at a depth of 3,400\,m w.e.  One important by-product of the 
operation of the prototype WARP detector was the precise determination of the $^{39}$Ar 
specific activity in natural argon.\\
$^{39}$Ar and $^{85}$Kr are two radioactive nuclides whose activity
in the atmosphere is of the order of 10\,mBq/m$^{3}$ and 
1\,Bq/m$^3$, respectively~\cite{loosli,kr85}.
As a result of the liquid argon production process, they are both present 
in abundant quantities and are the two most significant radioactive 
contaminations in liquid argon.
The two isotopes decay primarily by $\beta$ emission, and their presence can limit 
the sensitivity of experiments looking for low energy rare events (WIMP Dark Matter 
interactions, neutrinoless double beta decay) using liquid argon either as a target or as 
a shielding material.\\
$^{85}$Kr is not a pure $\beta$ emitter, owing to the presence of a 0.43\% branching 
ratio for decay with $\beta$ emission on a metastable state of $^{85}$Rb, which then 
decays by emitting a $\gamma$-ray of energy 514~keV, with a half-life of 
1.01\,$\mu$s~\cite{betaspec}. 
The coincidence between the $\beta$ and $\gamma$ emitted in a fraction of the $^{85}$Kr 
decays may ease in some cases the task of determining experimentally the activity of 
$^{85}$Kr in low-background detectors.  The determination of the specific 
activity of $^{39}$Ar is intrinsically more challenging and not as widely discussed 
in the literature~\cite{loosli,ms}. 
A theoretical estimate is presented in Ref.~\cite{cenniniAr}. \\
In the last twenty years liquid argon technology has acquired great relevance for 
astroparticle physics applications. Several experimental techniques, employing liquid 
argon as sensitive medium, have been proposed especially for rare events 
detection~\cite{warp-proposal,proposalRubbia,xenon93,icarust600,lanndd,snolab,protonDecay,arDM}. 
For WIMP Dark Matter direct detection, the discrimination of nuclear recoils from the 
$\beta$-$\gamma$ induced background plays a crucial role.  The discrimination provided 
by the experimental technique must be sufficient to reduce the radioactive background 
below the very low interaction rates foreseeable for WIMP Dark Matter.  A precise 
determination of the intrinsic specific activity of $^{39}$Ar is therefore of 
significant interest for the design of WIMP Dark Matter detectors employing argon 
as a target.
\section{The 2.3-liter WARP detector} \label{sec2}
\begin{figure}[tb]
\begin{center}
\epsfig{file=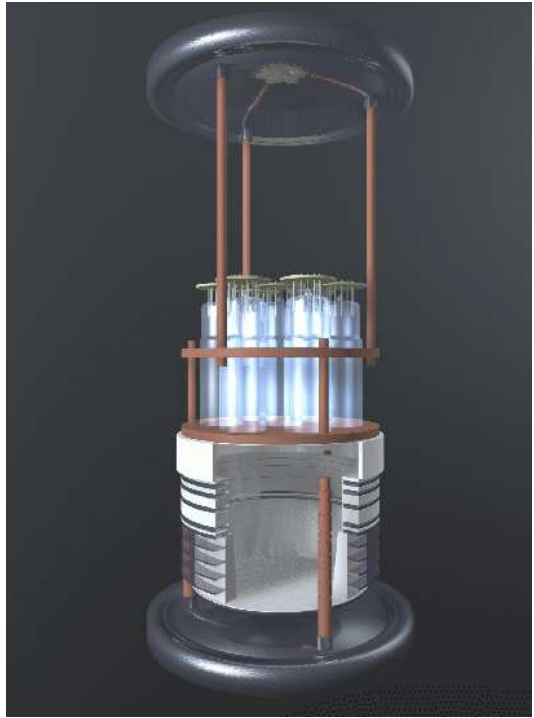, width=7cm}
\caption{Schematic view of the WARP 2 liter detector.}
\label{fig:warp25}
\end{center}
\end{figure}

The detector consists of a two-phase argon drift chamber with argon as a target. 
Two-phase argon drift chamber was first introduced within the ICARUS program~\cite{xenon93} 
in the framework of a wide-range study of the properties of noble gases.\\
The drift chamber (see Figure~\ref{fig:warp25}) 
is operated at the argon boiling point (86.7\,K) at the atmospheric pressure 
of the Gran Sasso Laboratory (about 950 mbar)~\cite{nist}. The cooling is provided by a 
thermal bath of liquid argon, contained 
in an open stainless steel dewar, in which the chamber is fully immersed.  The pressure of 
the gas phase on the top of the chamber is naturally equalized to the surrounding 
atmospheric pressure.\\
%%%%%%%% figure 1 %%%%%%%%%%%%%%%%
%%%%%%%%%%%%%%%%%%%%%%%%%%%%%%%%%%%%
Ionizing events inside the liquid argon volume produce scintillation VUV light mainly 
at 128\,nm. 
The scintillation light is shifted to wavelengths in the blue by an organic wavelength 
shifter (TetraPhenilButadiene) covering the walls, and collected by 
photomultipler tubes (PMTs) 
located in the gas phase and facing the liquid volume below.
The 2-inch PMTs are manufactured by Electron Tubes Ltd (model D757UFLA) and have a 
special photocathode that ensures functionality down to liquid argon 
temperatures. The quantum efficiency at the emission wavelength of TPB is about 
18\%. The material of the PMTs has been selected for high radiopurity: 
according to the supplier's specifications, the total $\gamma$ activity above 100~keV 
is 0.2~Bq/PMT, dominated by the $^{232}$Th and $^{238}$U chains.\\
A series of field-shaping rings surrounding the liquid phase superimpose an electric field 
of 1\,kV/cm.  The electrons are drifted toward the anode (located atop the chamber) and 
then 
extracted from the liquid to the gaseous phase by a local extraction field provided by a 
couple of grids.  The electrons are linearly multiplied in the gas phase by a second, 
stronger local field.  The PMTs detect the primary scintillation light (directly produced 
by the ionizing event) and also the secondary scintillation light (produced by the 
electron multiplication process in the gas phase). 
The PMT signals are summed and sent to a multi-channel analyzer recording the pulse 
height spectrum. The liquid argon contained in the chamber is Argon 
6.0 supplied by Rivoira S.p.A. 
The liquid argon is successively purified from electronegative impurities down to an equivalent 
contamination of less than 0.1\,ppb of O$_{2}$ by using the chemical filter Hopkalit\texttrademark~from Air 
Liquid.   The purity from 
electronegative elements is actively maintained by means of continuous argon 
recirculation through the chemical filter.\\
The experimental set-up is located in the Laboratori Nazionali del Gran Sasso underground 
laboratory (3,400 m w.e. of rock coverage).  The flux of cosmic ray muons is suppressed 
by a factor of  $10^{6}$ with respect to the surface (residual flux 1.1\,$\mu$/(m$^2 \cdot$~h), 
average muon energy $320$\,GeV~\cite{macro}).  The detector is shielded by 
10\,cm of lead, to reduce the external $\gamma$ background.  \\ 
The sensitive volume of the detector has the shape of frustum of cone and it 
is delimited by 
a stainless-steel cathode. The sensitive volume for the configuration under analysis is 
1.86$\pm$0.07\,liter. The density of liquid argon in the 
operating conditions (950\,mbar and 86.7\,K) 
is\,1.399 g/cm$^{3}$~\cite{nist}. 
The sensitive volume is viewed by seven PMTs, whose responses have been 
equalized in gain. Daily calibrations ensure the long-term stability and the linearity of the 
response. The sensitive volume and the argon thermal bath are contained in a stainless steel dewar, 
50~cm internal diameter and 200~cm internal height.
\section{Data analysis} \label{sec3}
\begin{figure}[tb]
\begin{center}
\begin{tabular}{@{}cc}
\epsfig{file=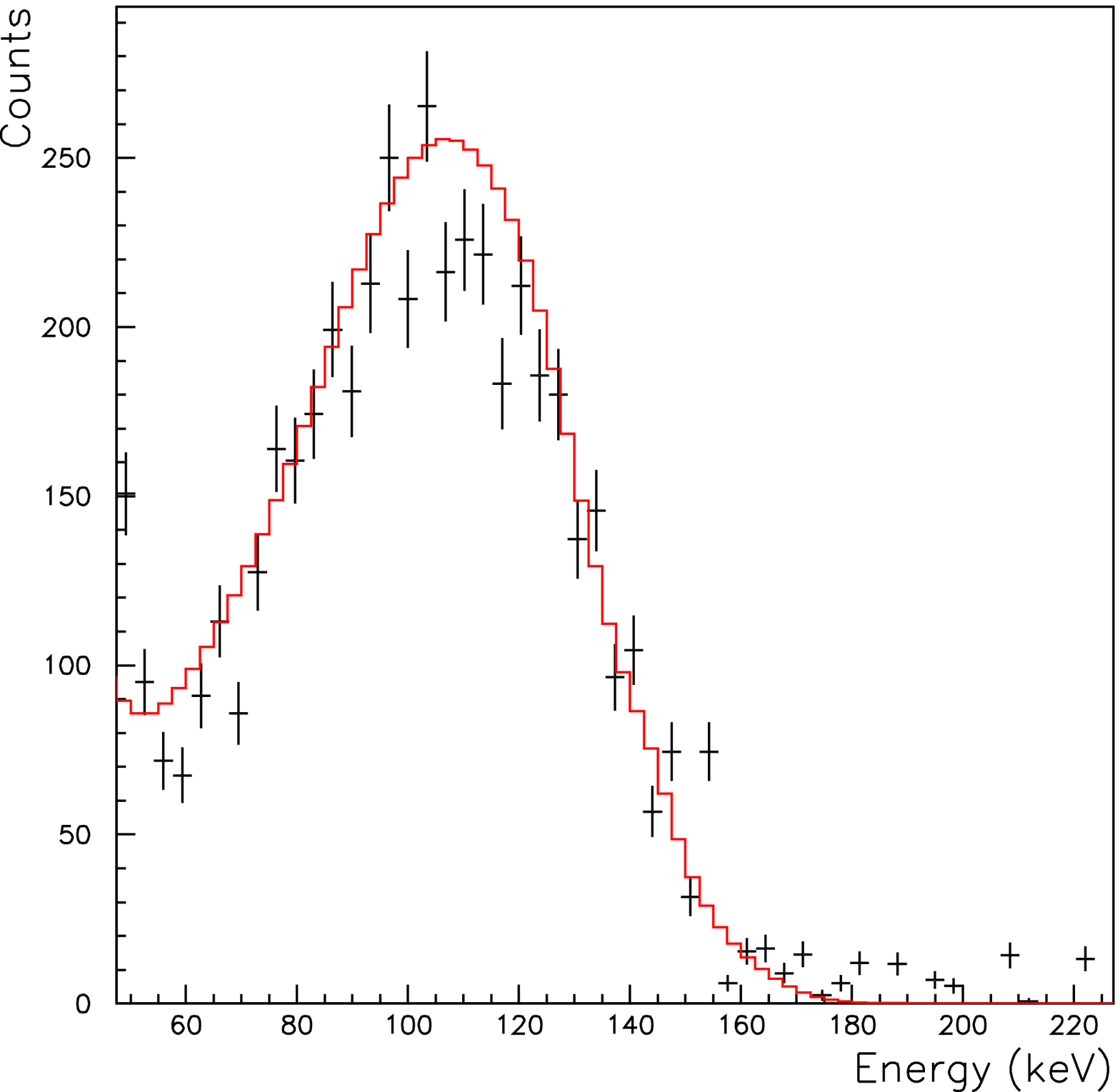, width=6.5cm}
(a)    &     \epsfig{file=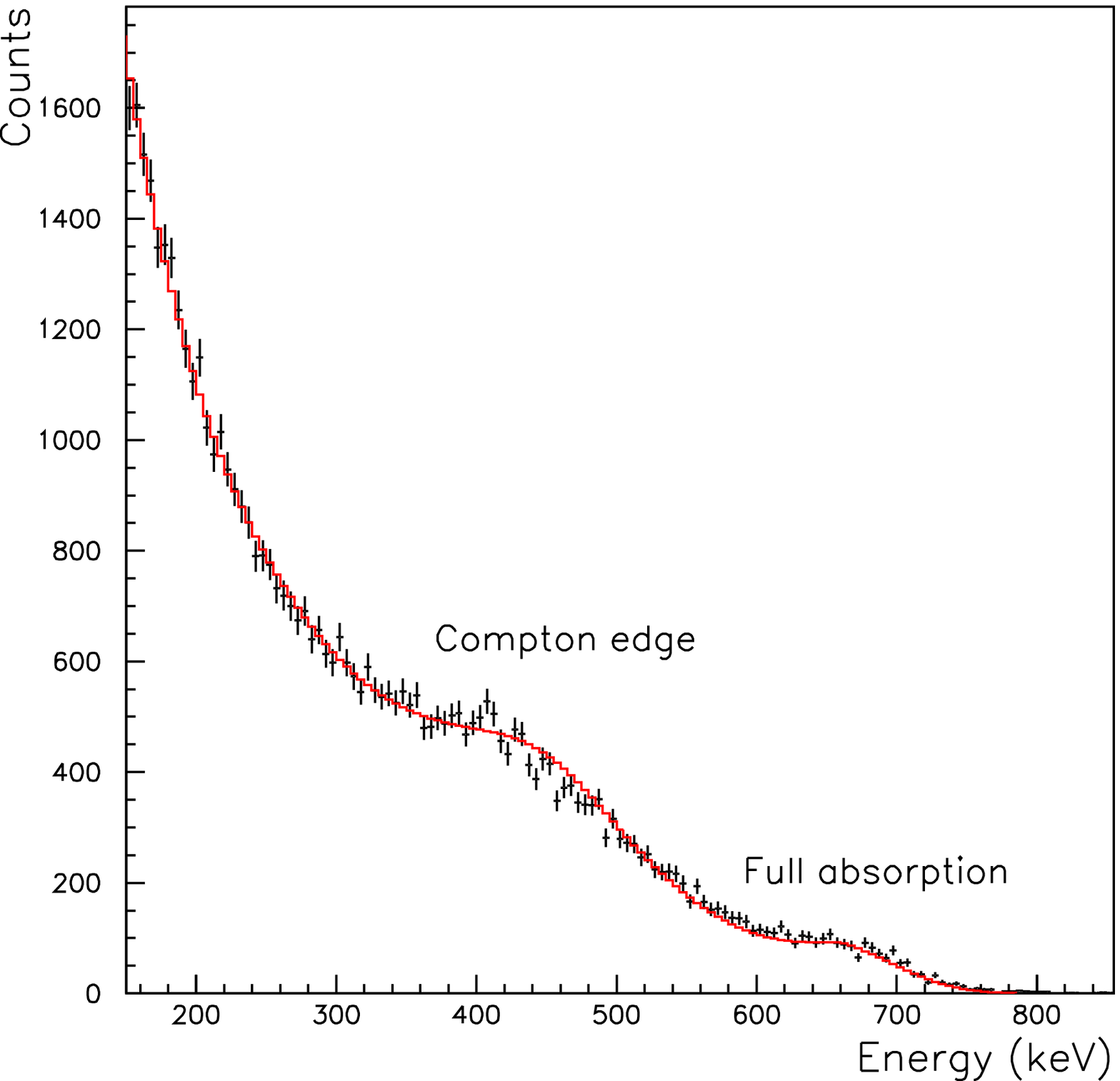, width=6.5cm} 
(b)   \\
\end{tabular}\\[2pt]
\end{center}
\caption{Energy spectra taken with external $\gamma$-ray sources, superimposed 
with the corresponding Monte Carlo simulations. 
(a) $^{57}$Co source ($E=122$~keV, B.R. 85.6$\%$,  and 136 keV, B.R. 10.7$\%$), 
(b) $^{137}$Cs source ($E=662$~keV).}
\label{fig:sources}
\end{figure}

\begin{figure}[htb]
\begin{center}
\epsfig{file=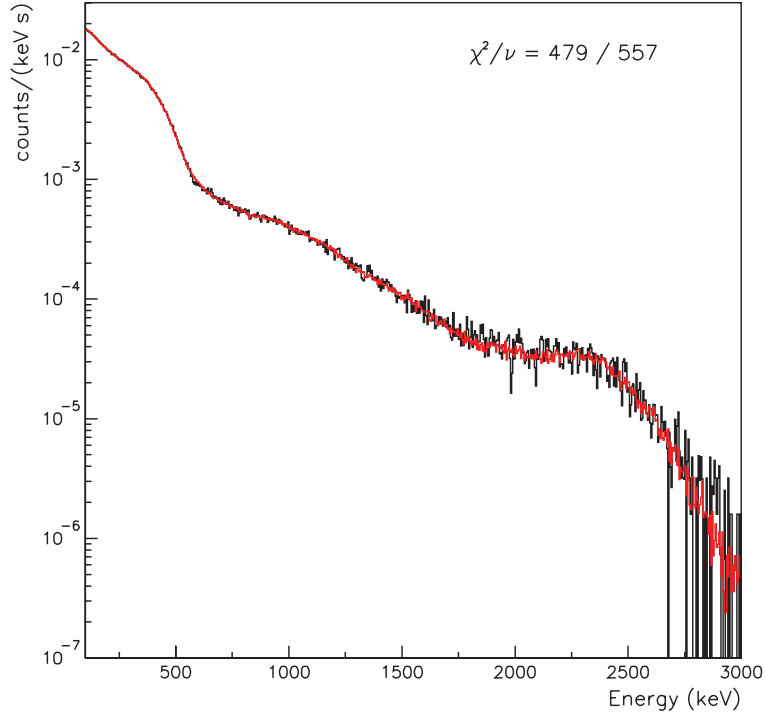, width=10cm}
\end{center}
\caption{Black histogram: background energy spectrum observed running the 2 liters detector
deep underground at LNGS inside a 10 cm thick Pb shielding. 
The superimposed red histogram is
the result of a fit with a Monte Carlo simulated signal (see text).}
\label{fig:spectrumtot}
\end{figure}

\begin{figure}[htb]
\begin{center}
\epsfig{file=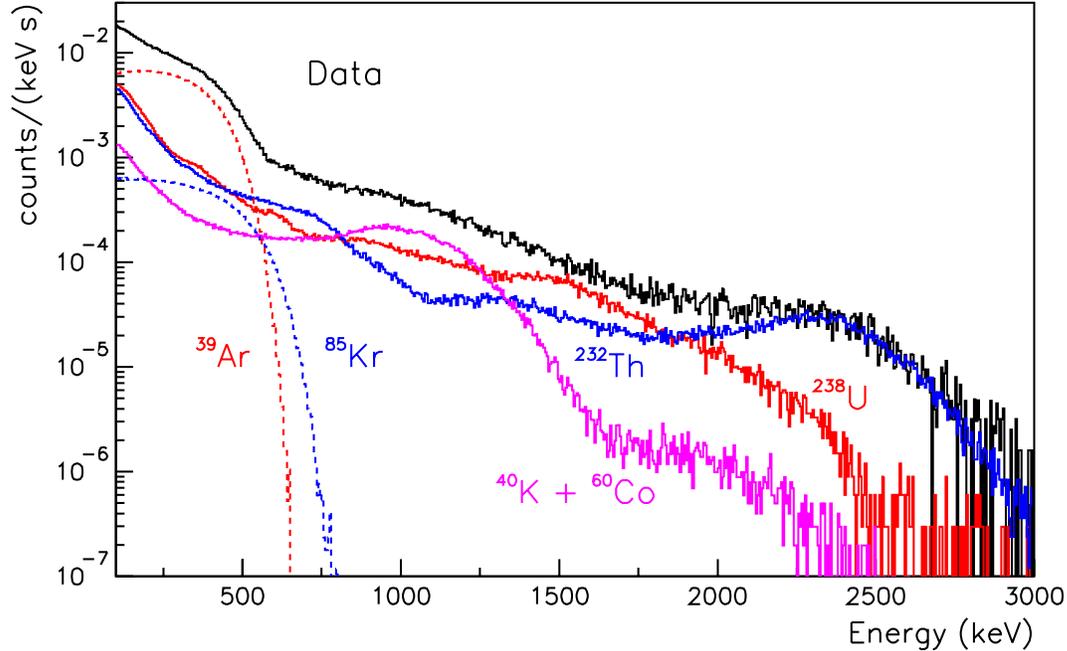, width=14cm}
\end{center}
\caption{The energy spectrum observed with the 2 liter
detector can be reproduced by: (a) an external component dominated
by interactions of $\gamma$-rays coming from U, Th, $^{60}$Co and 
$^{40}$K radioactivity of the materials surrounding the liquid argon; (b) an 
internal component dominated by  $^{39}$Ar and $^{85}$Kr $\beta$ contaminations 
inside the liquid argon.}\label{fig:sim}
\end{figure}

For the measurements described in this work, the electric fields were   
switched off and the chamber was operated as a pure scintillation detector. 
The gain of the 
PMTs has been set up to optimize the data acquisition in the typical energy range of the 
environmental $\gamma$-ray background, namely up to 3~MeV. The energy threshold for 
data acquisition is about 40~keV. The threshold used for analysis is 100~keV, in order 
to exclude events from electronic noise.
The response of the detector to $\gamma$ radiation was studied
using different $\gamma$-ray sources ($^{57}$Co, $^{60}$Co, $^{137}$Cs) 
placed outside the chamber. 
The spectra obtained with the $^{57}$Co and $^{137}$Cs sources are shown 
in Figure~\ref{fig:sources}. Typical values of the resolution 
observed with the calibration sources are $\sigma(E)/E = 13\%$ at 122~keV ($^{57}$Co) 
and $\sigma(E)/E = 6\%$ at 662~keV ($^{137}$Cs).
The correlation between energy and primary scintillation light detected 
was linear within the range tested with our sources.
The energy resolution of the detector can be described empirically
by the following parametrization:
\begin{equation}
\sigma (E) \ = \  \sqrt{a_{0}^{2} +a_{1}E+ (a_{2} E)^{2}}, \label{resolution}
\end{equation}
where $a_0 = 9.5$~keV, $a_1 = 1.2$~keV and $a_2 = 0.04$.
The three terms take into account the effects from non-uniform light collection
($a_{2}$ term), statistical fluctuations in the light production ($a_{1}$
term) and electronic noise ($a_{0}$ term). \\
%
%%%%%%%% figure 2 %%%%%%%%%%%%%%%%
The $\beta$-$\gamma$ spectrum in the detector 
(35 hours of live time) is displayed in Figure~\ref{fig:spectrumtot}. 
The total counting rate is about 6 Hz (4.2 Hz above the analysis threshold).  
A simulation based on the \textsc{Geant4} toolkit~\cite{geant4}  
has been developed to reproduce and understand the observed features. 
The simulation takes into account the following two components of background:
\begin{itemize}
\item 
External radiation from radioactive contaminants in the materials surrounding 
the liquid argon volume (stainless steel, thermal bath, PMTs). Only $\gamma$-emitters 
are taken into account, that originate an effective $\gamma$-ray flux through the 
surface of the sensitive volume. This includes: $^{238}$U and daughters 
(especially $^{222}$Rn dissolved in the external liquid argon), $^{232}$Th and 
daughters, $^{60}$Co and $^{40}$K. The presence of these sources has been 
confirmed with a portable NaI $\gamma$-spectrometer inserted into the empty dewar.
\item Bulk contaminations in the liquid argon of 
the chamber. In this case, both $\beta$ and $\gamma$ emitters are relevant. 
The most important contributions 
are from $^{222}$Rn, $^{39}$Ar and $^{85}$Kr.\footnote{Other radioactive isotopes 
of Ar, as $^{37}$Ar and $^{41}$Ar, are short-lived ($T_{1/2}$ are 35 days 
and 109 min, respectively) and their cosmogenic production is negligible 
in the underground 
laboratory. The long-lived $^{42}$Ar ($T_{1/2}$ = 32.9~y) is 
expected to be present in natural argon because of thermonuclear tests 
in the atmosphere; however, its concentration in $^{\rm nat}$Ar is
negligible, $< 6 \cdot 10^{-21}$ 
g/g (at 90\% CL)~\cite{ar42}, corresponding to less than 85 
$\mu$Bq/liter in liquid argon.}
\end{itemize}
The signal from internal $^{222}$Rn and its daughters can be monitored by counting 
the $\alpha$ decays of the isotopes $^{222}$Rn, $^{218}$Po and $^{214}$Po 
in the energy region 5$-$7 MeV.
After a new filling with freshly produced liquid argon the chamber shows 
a total $^{222}$Rn decay rate of the order of 1$-$2 Hz. 
Due to the $^{222}$Rn half-life of 3.8 days, 
the observed rate decreases to few tens of events/day four weeks after 
the filling. Therefore, $\beta$ decays from the Rn daughters 
$^{214}$Pb and $^{214}$Bi can be neglected, provided the measurement 
is performed a few weeks after the filling of the chamber. The counting 
rate due to the decay of $^{14}$C (dissolved in the liquid 
argon or located in the surrounding plastics) is estimated to be 
much less than 50 mHz. Most of the $^{14}$C events occur close 
to the chamber walls.\\
The main characteristics of the $^{39}$Ar and $^{85}$Kr $\beta$ decays
are summarized in Table~\ref{table:1}. Since both decays are classified as 
forbidden unique $\beta$ transitions ($\Delta I^{\Delta \pi} = 2^{-}$), the 
$\beta$ spectrum is not described by the usual Fermi function. For the 
present work, we assumed the $\beta$ spectra from Ref.~\cite{betaspec}. \\
\begin{table*}[ttt]
\caption{Characteristics of the $\beta$ decays of 
$^{39}$Ar and $^{85}$Kr.}
\label{table:1}
\newcommand{\m}{\hphantom{$-$}}
\newcommand{\cc}[1]{\multicolumn{1}{c}{#1}}
\renewcommand{\tabcolsep}{2pc} % enlarge column spacing
\renewcommand{\arraystretch}{1.2} % enlarge line spacing
\begin{tabular}{@{}lccc}
\hline
Isotope           & Half-life & $\beta$ end-point & $\beta$ mean energy  \\
                  &           &    (keV)          &    (keV)      \\
\hline
$^{39}$Ar         & 269 y   & 565             & 220\\
$^{85}$Kr         & 10.8 y  & 687            & 251 \\
\hline
\end{tabular}\\[2pt]
\end{table*}
The \textsc{Geant4}-based simulation is used to generate 
normalized spectra $s_i(E)$ from the  
different radioisotopes, taking into account the energy 
resolution of the detector. The isotopes from the natural radioactive 
chains are treated independently.    
A $\chi^2$ fit of the experimental spectrum $F(E)$ is then performed 
in the energy range from 100 keV to 3 MeV with a linear 
combination of the single components, i.e. 
\begin{equation}
F(E) = \sum_i{w_i \cdot s_i(E)}.
\end{equation}
The coefficients $w_i$ are treated as free parameters and 
represent the counting rates induced by the single sources. 
In Figure~\ref{fig:spectrumtot} we show the experimental spectrum,  
superimposed with the output of the fit (i.e. $\sum w_{i} s_{i}(E)$). 
The fit is satisfactory in all the energy range 
considered. % ($\chi^2$=551 with 545 d.o.f.).
The signals from the most important external $\gamma$-rays radioactivity sources 
and from internal contaminations are shown in Figure~\ref{fig:sim}, as derived 
from the analysis of the experimental spectrum. 
%\textcolor{red}{The fit results are stable if 
%the energy range of the fit is changed or if secular equilibrium is assumed.}
%
%%%%%%%% figure 3 %%%%%%%%%%%%%%%%
%
%%%%%%%% figure 4 %%%%%%%%%%%%%%%%
%
%%%%%%%% figure 5 %%%%%%%%%%%%%%%%
\begin{figure}[htb]
\begin{center}
\epsfig{file=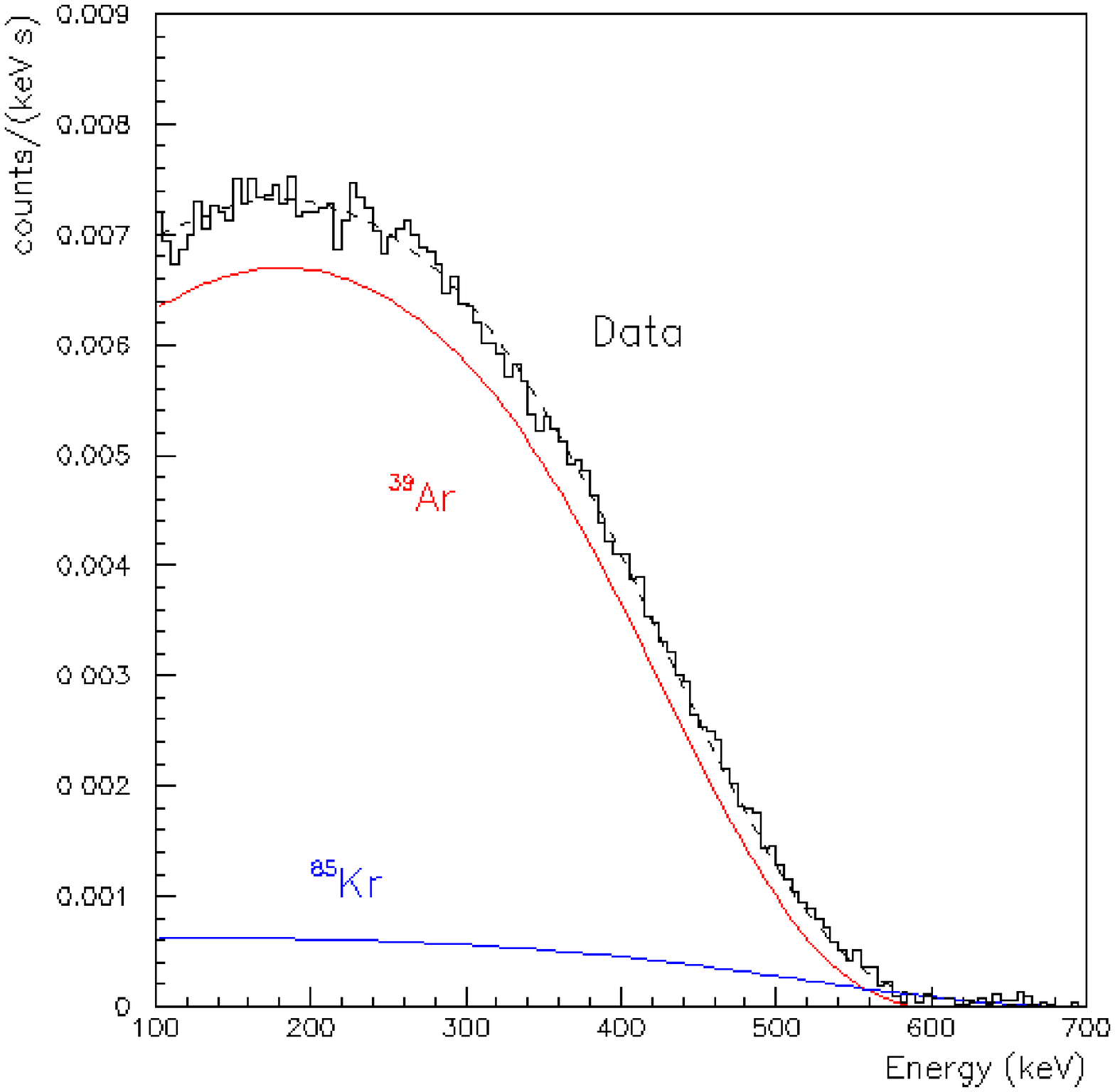, width=10cm}
\end{center}
\caption{Residual energy spectrum observed in the WARP chamber after subtraction
of the expected contribution from external $\gamma$-rays. The spectrum is well described 
by a combination (dashed line) of $\beta$ decays of 
$^{39}$Ar and $^{85}$Kr.}\label{fig:spectrumar}
\end{figure}
\section{Discussion}\label{sec4}
\begin{figure}[htb]
\vspace{9pt}
\begin{center}
\epsfig{file=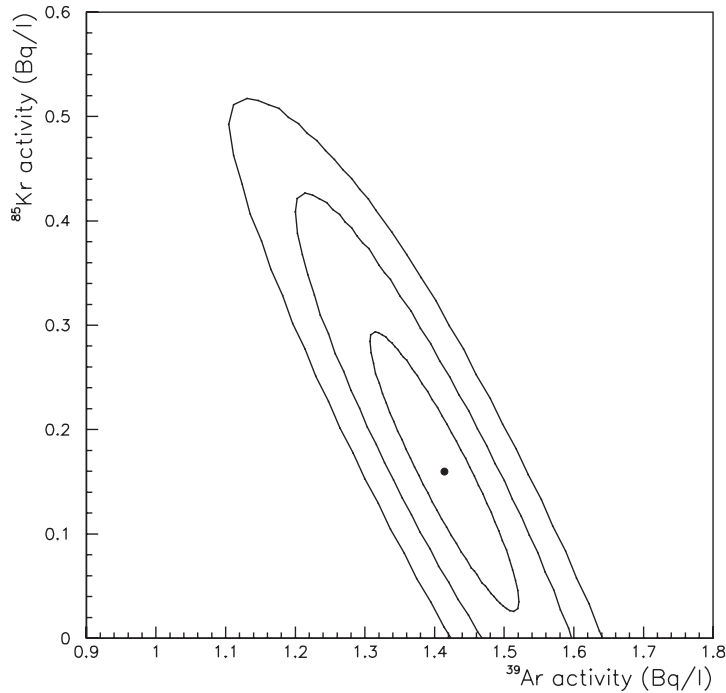, width=9.5cm}
\end{center}
\caption{Confidence regions ($1 \sigma$, $2 \sigma$ and $3 \sigma$) for the $^{39}$Ar and 
$^{85}$Kr specific activities. Systematic uncertainties have been added in quadrature.}
\label{fig:corr}
\end{figure}

Figure~\ref{fig:sim} shows that the energy region 2$-$3 
MeV is dominated by interactions of $\gamma$-rays from $^{232}$Th daughters, the
region 1.5$-$2 MeV by $\gamma$-rays from $^{238}$U daughters, and the region 0.5$-$1.5 
MeV by $\gamma$-rays from $^{60}$Co and $^{40}$K. 
Below 0.5 MeV the main contribution comes from $\beta$ decays from
internal contaminations of $^{39}$Ar and $^{85}$Kr; the 
two isotopes account for 65\% of the total counting rate between 100 
and 500~keV. \\
The cosmogenically-originated $^{39}$Ar contamination of $^{\rm nat}$Ar 
in the troposphere was measured in Ref.~\cite{loosli} to be 
$(7.9 \pm 0.3) \cdot 10^{-16}$ g/g; the quoted error was statistical 
only\footnote{The $^{39}$Ar/$^{\rm nat}$Ar ratio was measured for dating purposes.  
The knowledge of the absolute $^{39}$Ar specific activity was hence not necessary.}. 
Since the liquid argon used for the experiment is produced from the atmospheric 
gas, a similar $^{39}$Ar/$^{\rm nat}$Ar ratio is expected to be present in our sample. \\
$^{85}$Kr is mainly produced as fission product of uranium and plutonium. Its abundance in the 
atmosphere is of the order of 1 Bq/m$^{3}$, corresponding to about
$4 \cdot 10^{-15}$ g($^{85}$Kr)/g($^{\rm nat}$Ar) in air. 
Nevertheless, the distillation procedure for the production of liquid argon substantially 
reduces the $^{85}$Kr fraction. The residual $^{85}$Kr in liquid argon 
may vary in different batches of liquid. \\ 
In order to better show the Ar and Kr signals, 
Figure~\ref{fig:spectrumar} displays the spectrum obtained from the experimental
data after subtracting the fitted contribution from the other sources. 
The single $^{39}$Ar and $^{85}$Kr contributions can be  
disentangled from the different end-point energies, 565 keV and 687 keV respectively. 
Since $^{85}$Kr and $^{39}$Ar decays populate the same energy region of the spectrum 
their estimates are anti-correlated, as displayed in Figure~\ref{fig:corr}. \\
The specific activity of $^{39}$Ar in liquid argon resulting from the analysis is 
1.41$\pm$0.02(stat)$\pm$0.11(syst) Bq/liter ($1 \sigma$) (see below for the discussion 
of systematic uncertainties). The corresponding $^{39}$Ar/$^{\rm nat}$Ar mass ratio 
is $(8.0 \pm 0.6) \cdot 10^{-16}$ g/g, with errors summed in quadrature. 
The result is in excellent agreement with the atmospheric determination 
of Ref.~\cite{loosli}. \\
From the fit it is also obtained that the $^{85}$Kr activity in the sample under 
investigation is (0.16$\pm$0.13) Bq/liter ($1 \sigma$). In a previous measurement 
performed with a sample of Argon 99.999\% from Air Liquid, it was found a $^{85}$Kr 
activity three times larger. This indicates that a non-negligible $^{85}$Kr 
contamination may be found in commercial liquid argon samples.  
In this case, an additional fractional distillation could be required to 
reduce the radioactive background for the WARP experiment.\\
\begin{table*}[tb]
\caption{Systematic uncertainties on the $^{39}$Ar specific 
activity.}
\label{table:2}
\newcommand{\m}{\hphantom{$-$}}
\newcommand{\cc}[1]{\multicolumn{1}{c}{#1}}
\renewcommand{\tabcolsep}{2pc} % enlarge column spacing
\renewcommand{\arraystretch}{1.2} % enlarge line spacing
\begin{tabular}{@{}llll}
\hline
Item & Relative error & Absolute error (Bq/liter) \\
\hline
Energy calibration & $\pm$6.5\% & $\pm$0.092  \\
Energy resolution & $\pm$1.3\% & $\pm$0.018 \\
Sensitive mass & $\pm$3.8\% & $\pm$0.054 \\
\hline
Total & & $\pm$ 0.11 \\
\hline
\end{tabular}\\[2pt]
\end{table*}
The systematic uncertainties are summarized in Table~\ref{table:2}.
The dominating item is related to the energy calibration of the detector 
response: since the discrimination between $^{39}$Ar and $^{85}$Kr is based 
upon the $\beta$ end-points, the fit result is sensitive to the energy 
calibration and resolution. The uncertainty on the energy calibration in the range 
of interest was evaluated to be 2\% ($1 \sigma$) from the meaurements with
the $\gamma$-ray sources; the corresponding $^{39}$Ar systematic 
error is 6.5\%. 
The second important contribution is related to the uncertainty in the
active volume of the chamber. 
The filling level can be determined with accuracy of about 1 mm, and the
diameter of the teflon container with the reflector fixed on it is 
known with a precision of about 2~mm. This corresponds to an uncertainty on the
sensitive mass of 3.8\%.
%
%%%%%%%% figure 6 %%%%%%%%%%%%%%%%
\section{Conclusions}\label{sec5}
The best estimate of the $^{39}$Ar specific activity in the liquid argon 
is $(1.41 \pm 0.11)$ Bq/liter, or $(1.01 \pm 0.08)$ Bq/kg of natural Ar, or 
$(8.0 \pm 0.6) \cdot 10^{-16}$ g($^{39}$Ar)/g($^{\rm nat}$Ar). The value is  
consistent with the previous determination by H.~Loosli~\cite{loosli}. 
The uncertainty in our measurement is mainly due to systematics. \\ 
The liquid argon sample under investigation shows a contamination 
of $^{85}$Kr, 0.16$\pm$0.13 Bq/liter ($1 \sigma$).
\section{Acknowledgments} \label{ack}
We wish to dedicate this work to the memory of our friend and colleague 
Nicola Ferrari, co-author of the paper, who prematurely passed away on 
July, 2006. \\ 
We also thank Prof.~H.~Loosli for helpful communications concerning 
his $^{39}$Ar paper. L.~P. acknowledges the support by the EU FP6 
project \textsc{Ilias}. A.~M.~S. has been in part supported by a grant 
of the President of the Polish Academy of Sciences, by the \textsc{Ilias} 
contract Nr.~RII3-CT-2004-506222 and by the MNiSW grant 1P03B04130.

\end{document}